\documentclass[epj,nopacs]{svjour}
\usepackage{graphics}
\usepackage{makecell}
\usepackage[numbers]{natbib}
\usepackage{notoccite}

\usepackage{graphics}
\usepackage{makecell}
\usepackage[numbers]{natbib}
\usepackage{notoccite}

\usepackage{gensymb}
\usepackage{dcolumn}
\usepackage{float}
\usepackage{mathptmx, courier, pifont}
\usepackage[scaled=0.92]{helvet}
\usepackage[T1]{fontenc}
\usepackage{textcomp}
\usepackage{color}
\usepackage[normalem]{ulem}

\usepackage[dvipsnames]{xcolor}
\usepackage[colorlinks=true,urlcolor=Blue,linkcolor=Blue]{hyperref}
\usepackage[all]{hypcap}
\usepackage{makecell}
\usepackage{booktabs}
\usepackage{multirow}

\begin{document}
\title{High-precision measurement of the atomic mass of $^{84}$Sr and implications to isotope shift studies}
\author{Z.~Ge\inst{1}\thanks{\emph{Corresponding author:} zhuang.z.ge@jyu.fi} \and 
{S.~W.~Bai\inst{2}}\and
{T.~Eronen\inst{1}}\and 
{A.~Jokinen\inst{1}}\and
{A.~Kankainen\inst{1}}\and
{S.~Kujanp\"a\"a\inst{1}}\and
{I.~D.~Moore\inst{1}}\and
{D.~A.~Nesterenko\inst{1}}\and
{M. Reponen\inst{1}}
}                     
%
%
\institute{{Department of Physics, University of Jyv\"askyl\"a, P.O. Box 35, FI-40014, Jyv\"askyl\"a, Finland} \and {School of Physics and State Key Laboratory of Nuclear Physics and Technology, Peking University, Beijing 100871, China}
}
\date{\today}
%
\abstract{
The absolute mass of $^{84}$Sr was determined using the  phase-imaging ion-cyclotron-resonance technique  with the JYFLTRAP double Penning trap mass spectrometer. A more precise value for the mass of $^{84}$Sr is essential for providing potential indications of physics beyond the Standard Model through high-precision isotope shift measurements of Sr atomic transition frequencies. The  mass excess of $^{84}$Sr was refined to be -80649.229(37) keV/c$^2$ from high-precision cyclotron-frequency-ratio measurements with a relative precision of 4.8$\times$10$^{-10}$. The obtained mass-excess value is in agreement with the adopted value in the  Atomic Mass Evaluation 2020, but is 30 times more precise. 
With this new value, we confirm the previously observed nonlinearity in the study of the isotope shift of strontium. Moreover, the double-beta ($2\beta^{+}$) decay $Q$ value of $^{84}$Sr was directly determined to be 1790.115(37) keV, and the precision was improved by a factor of 30.
\keywords { {Penning trap}}--{High precision mass spectrometry}--{isotope shift}--{Double beta  decay $Q$ value}
%
} 
\maketitle
\section{Introduction}


Isotope shifts in atomic transition frequencies arise from differences in neutron numbers 
between isotopes
sharing the same atomic number. 
The isotopic shifts in the frequency of an atomic transition 
show
an approximately linear correlation with the isotope shift observed in a second transition. These shifts reveal contributions from field and mass shifts~\cite{king1984isotope}, originating from the differing nuclear masses of isotopes and variations in their nuclear charge distribution. 
Isotopic shifts can be systematically studied using a King plot analysis. In this analysis, the isotope shifts 
in two electronic transitions
within the same isotopes are correlated. To perform the King plot analysis, one measures the energies of two transitions for three or more isotopes of a specific element.
The King plot is expected to exhibit linearity~\cite{king1984isotope,Athanasakis23}, with the experimentally determined slope serving as a reliable benchmark for theoretical predictions~\cite{flam18}. Deviations from linearity are crucial for refining atomic structure calculations~\cite{dammalapati2009,shi2017,naze2015}. 
Recent theoretical proposals suggest that the observed nonlinearity in King plots could be 
used
to impose constraints on higher-order effects on field isotope shift within the Standard Model (SM) or  a possibility of a new interaction mediated by a boson beyond SM~\cite{frugiuele2017,Berengut2018,Flambaum18}.
SM contributions to field isotope shift, which include quadratic field shift, relativistic effects, and effects of nuclear deformation, are proposed to account for nonlinearities of the King plot~\cite{Munro-Laylim22}. 
In Refs. ~\cite{Flambaum18,Flambaum21}, nuclear polarization has also been considered as a non-negligible contribution to the nonlinearity.

Motivated by this, a recent surge in efforts ~\cite{miyake2019,manovitz2019,knollmann2019,counts2020,solaro2020,NESTERENKO2020-2,Counts20,rehbehn2021,muller21,hur2022,figueroa2022,ono2022} has significantly advanced the precision of isotope shift measurements. 
Essential to placing constraints on proposed electron-neutron interactions and other novel physics through King's linearity is a two-fold requirement. First, experimental data in the form of precision optical spectroscopy and 
atomic mass measurements
are needed to empirically constrain the potential size of the nonlinearity. Second, if a nonlinearity is observed, precise atomic and nuclear theory is necessary to calculate beyond-first-order SM sources of nonlinearity~\cite{frugiuele2017,Berengut2018,Flambaum18,Delaunay17}. 
A natural inquiry emerges regarding the potential for the King plot to maintain its linearity at an enhanced level of experimental precision. Recent experiments conducted with strontium and ytterbium ions ~\cite{miyake2019,Counts20} have provided initial indications that this linearity is, in fact, disrupted at a magnitude of several standard deviations.
Strontium presents favorable properties for studying isotope shifts, boasting an abundance of stable isotopes and very narrow optical transitions~\cite{stellmer2014}. Earlier theoretical work has also proposed the measurement of strontium isotope shifts as a promising probe for new physics~\cite{frugiuele2017,Berengut2018,Berengut2020}. Alkaline-earth element strontium features four stable isotopes: three bosons ($^{84,86,88}$Sr) and one fermion ($^{87}$Sr). 
Precise and accurate determination of the atomic mass of the stable Sr isotopes is crucial for probing potential causes for such nonlinearities.

%
\begin{figure*}[!htb]
\centering
\resizebox{0.999\textwidth}{!}{%
  \includegraphics{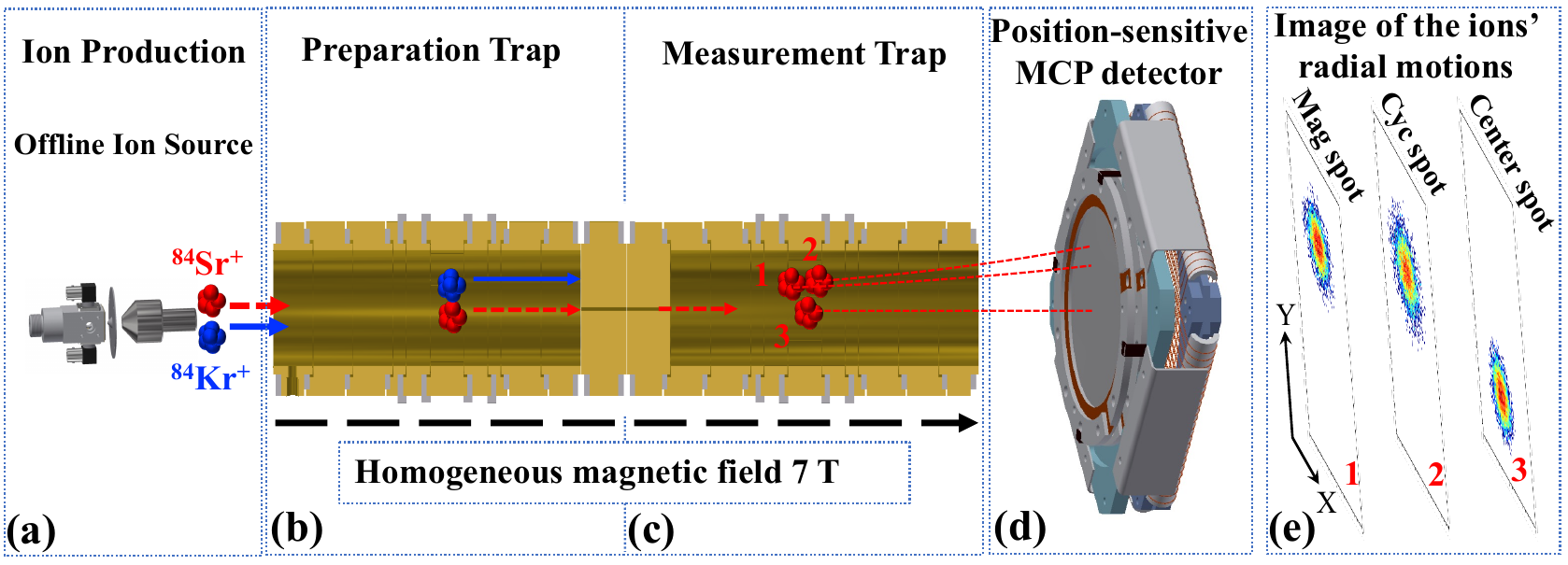}
}
\caption{(Color online).  Schematic view of ion production and mass measurements using the PI-ICR technique at IGISOL. 
Stable $^{84}$Kr$^+$ and $^{84}$Sr$^+$ ions were generated using an offline glow-discharge ion source (a). 
Ions having mass number of 
84
were selected with a dipole magnet  and transported to the JYFLTRAP PTMS for final ion species selection in the  preparation trap (b) by means of a buffer-gas cooling technique and cyclotron frequency determination using the phase-imaging technique in the measurement trap (c). A position-sensitive MCP detector (d) was used to register the images of the motion  phases. (e) An illustration of the radial-motion (``magnetron'', ``cyclotron'', and ``center'') projection of the $^{84}$Sr$^+$ ions onto the position-sensitive MCP detector. Each pixel’s color corresponds to a different number of ions.
}
\label{fig:igisol}    
\end{figure*}
%

Atomic masses of $^{86-88}$Sr  have been measured at the FSU Penning trap ~\cite{rana2012}  with high precision (uncertainty  of $\leq$ 6 eV/c$^{2}$) as adopted in the most recent Atomic Mass Evaluation 2020 (AME2020)~\cite{Wang2021,Kondev2021}. 
The atomic mass uncertainty for $^{84}$Sr, the least abundant naturally occurring isotope of strontium, is however notably higher at 1.2 keV/c$^2$. For investigating fundamental physics and exploring phenomena beyond the SM via high-precision King-plot tests with strontium, it is critical to measure the atomic mass of $^{84}$Sr directly with high precision. In this article, we report on the most precise absolute mass value of $^{84}$Sr to date, achieved via high-precision cyclotron-frequency-ratio measurements with the JYFLTRAP double Penning trap mass spectrometer (PTMS). We employ this refined mass value in an updated King plot analysis for two transitions of strontium.
\section{Experimental method}
The experiment was performed at the Ion Guide Isotope Separator On-Line facility (IGISOL) using the JYFLTRAP double PTMS~\cite{Eronen2012a,Eronen2016}, at the University of Jyv\"askyl\"a~\cite{Moore2013}.  
Stable $^{84}$Sr$^+$ ions were generated using an offline glow-discharge ion source~\cite{Vilen2020a}. For a precise mass measurement of $^{84}$Sr, reference ions of $^{84}$Kr$^+$ with a well-known mass value (uncertainty of 4 eV/c$^{2}$~\cite{Wang2021}) were concurrently produced from the same ion source. As illustrated in Fig.~\ref{fig:igisol}(a), the gas cell comprising the glow-discharge ion source hosts two sharp electrodes, with one composed of naturally abundant strontium. By introducing a gas containing natural krypton, it facilitates the simultaneous generation of stable ions for both strontium and krypton.

The generated ions of $^{84}$Sr$^+$ and $^{84}$Kr$^+$ were extracted using helium gas flow and electric fields facilitated by a sextupole ion guide (SPIG)~\cite{Karvonen2008}. Following acceleration over an electric potential of $\approx$30 kV, the ions with mass number of $A$ = 84 were mass-separated using a 55$^\circ$ dipole magnet with a mass resolving power of $M/\Delta{M}$ $\approx$ 500.
Post isobaric separation, the ion beam passed through a pulsed electrostatic kicker, which functioned as a beam gate to regulate the ion rate. 
The chopped ions, controlled by the beam gate, were directed to a radiofrequency-quadrupole cooler-buncher (RFQ-CB)~\cite{Nieminen2001}, where they were accumulated, cooled, and bunched.

The ion bunches from the RFQ-CB were injected to the JYFLTRAP double PTMS, consisting of two cylindrical Penning traps equipped with a 7-T superconducting solenoid. The first trap, functioning as a purification trap, is filled with buffer gas and is employed for isobaric purification using the sideband buffer-gas cooling technique~\cite{Savard1991}. This technique alone achieves mass purification with a resolving power of approximately $10^{5}$ by selectively converting ion motion from magnetron to reduced cyclotron motion.

In the purification trap, all cooled and centered ions ($^{84}$Sr$^{+}$ and $^{84}$Kr$^{+}$) were initially excited to a large orbit of revolution by applying a dipole excitation at the magnetron motion frequency $\nu_{-}$ for approximately 11 ms. Subsequently, a quadrupole excitation 
was applied at the cyclotron frequency of the ions of interest (only $^{84}$Sr$^{+}$ or $^{84}$Kr$^{+}$) for about 100 ms,  to center them through collisions with the buffer gas.
This technique alone can provide sufficient cleaning for $^{84}$Sr$^{+}$ or $^{84}$Kr$^{+}$. A even higher resolving power selecting method, the Ramsey cleaning technique, was additionally employed with a resolving power better than $10^6$~\cite{Eronen2008a} right after the sideband buffer-gas cooling to ensure no leaking events of contaminants. 
In this method, the ions extracted through a 1.5-mm diaphragm to the second trap (measurement trap) undergo an additional cleaning step utilizing a dipolar excitation with time-separated oscillatory fields at the mass-dependent reduced cyclotron frequency ($\nu_{+}$), which selectively increases the cyclotron radius of the contaminants. The contaminants were implanted on the diaphragm after subsequent transfer back to the first trap. A purified sample of either $^{84}$Sr$^{+}$ or $^{84}$Kr$^{+}$ ions was prepared after selection and cleaning using the aforementioned techniques. 
Finally, these ions were centered again in the first trap and transferred to the second trap for measuring the actual cyclotron frequency.



\begin{figure}[!tbh]
\centering
\resizebox{0.49\textwidth}{!}{%
  \includegraphics{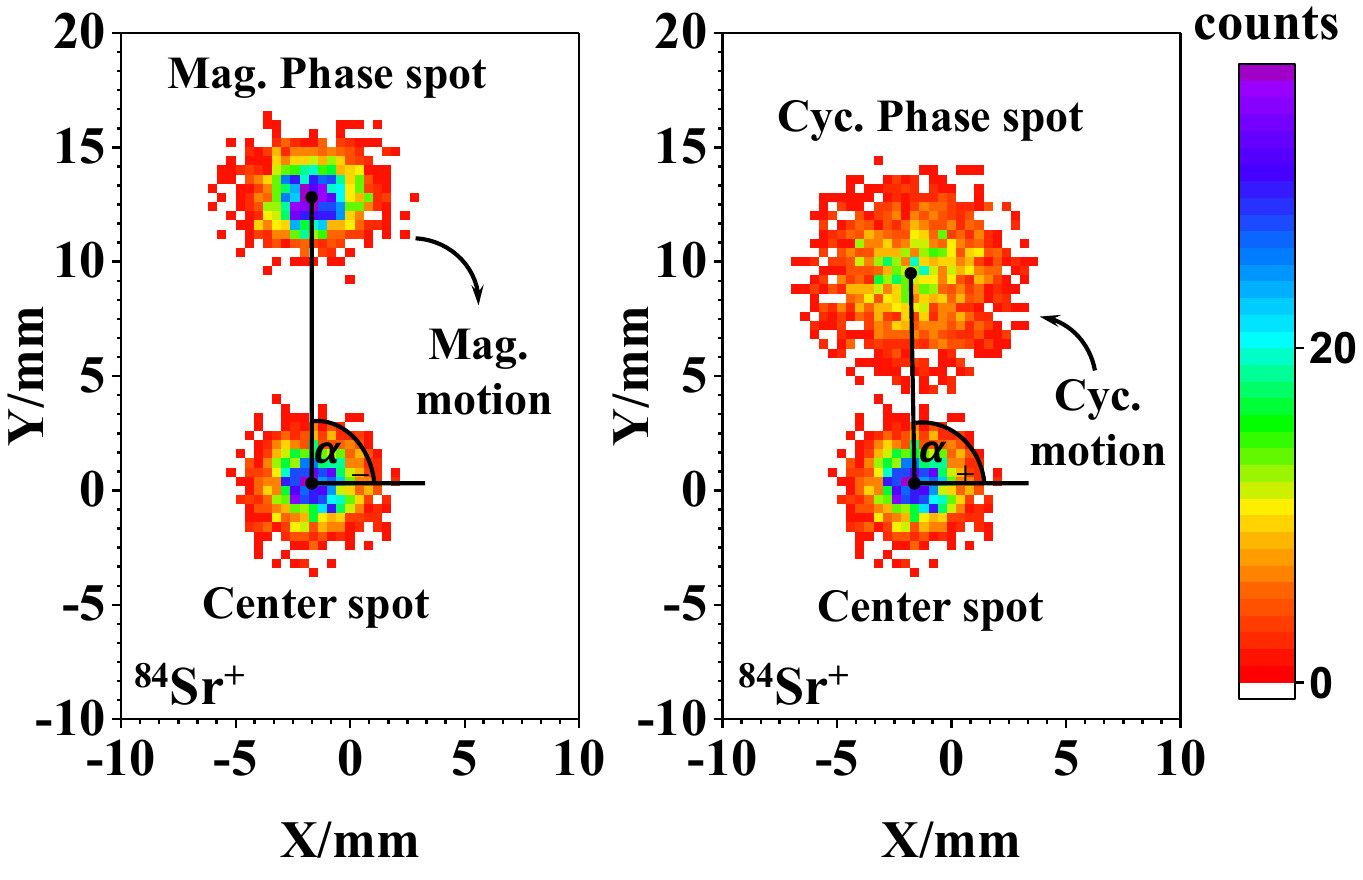}
  }
   \caption{(Color online). Ion spots (magnetron phase, cyclotron phase, and center) of $^{84}$Sr$^{+}$ on the 2-dimensional position-sensitive MCP detector after a typical PI-ICR excitation pattern with an accumulation time of 400 ms. The magnetron phase spot is shown on the left side  and the cyclotron phase spot on the right. 
   By analyzing the angle difference between the two phase spots relative to the center spot, $\alpha_c = \alpha_+ - \alpha_-$, the cyclotron frequency of the measured ion species can be deduced. Color bars indicate the number of ions in each pixel.
   }
   \label{fig:PI-ICR-2-phases}
\end{figure}

In the second trap, the phase-imaging ion-cyclotron-resonance (PI-ICR) method~\cite{Nesterenko2018,nesterenko2021} was employed to measure the cyclotron frequency, $\nu_{c}= {qB}/{(2\pi m)}$, 
where $B$ is the magnetic field strength, $q$ is the charge state, and $m$ is the mass of the stored ion.
The scheme of the PI-ICR technique~\cite{nesterenko2021,Nesterenko2018,Eliseev2014,Eliseev2013} at JYFLTRAP relies on direct measurements of the cyclotron motion and magnetron motion simultaneously by projecting the radial ion motion onto a position-sensitive MCP detector.
To determine the phases of the radial motions, the center has to be determined for the ion spots on the detector. This is done by storing the ions for a few milliseconds without exciting 
their cyclotron motion, after which the ions
are directly extracted from the trap and projected onto the MCP detector.
Two patterns, as detailed in~\cite{Nesterenko2018,nesterenko2021}, are utilized to measure the magnetron or cyclotron motion phases, respectively. The angle between two phase images of the projected radial motions with respect to the center spot is $\alpha_c = \alpha_+ - \alpha_-$, where $\alpha_+$ and $\alpha_-$ are the polar angles of the cyclotron and magnetron motion phases. The cyclotron frequency $\nu_{c}$ is derived from: 
\begin{equation}
\label{eq:vc}
\nu_{c}=\frac{\alpha_{c}+2\pi n_{c}}{2\pi{t_{acc}}}, 
\end{equation}
where $n_{c}$ is 
the sum of the numbers of full revolutions in the two patterns of the measured ions during the phase accumulation time $t_{acc}$. 
A few different accumulation times for $^{84}$Sr$^{+}$ and $^{84}$Kr$^{+}$ were used to confirm unambiguously the cyclotron frequency.
A fixed accumulation time of 400 ms was employed for the actual measurements to determine the final $\nu_{c}$. 
A measurement with "cyclotron" and "magnetron" phase spots collected with respect to the center spot is schematically shown in Fig.\ref{fig:igisol}(e). 
A representative measurement of the magnetron and cyclotron phase spots relative to the center spot is shown in the left and right panels of Fig.~\ref{fig:PI-ICR-2-phases}, respectively.

The atomic mass $M_{^{84}Sr}$ was derived from the measured cyclotron frequency ratio ($R$ = ${\nu_{c}(^{84}Kr^{+}})/{\nu_{c}}(^{84}Sr^{+})$) measurements  of singly charged ions of the decay pair $^{84}$Sr-$^{84}$Kr:
\begin{equation}
\label{eq:mass}
M{(^{84}Sr)} = R(M{(^{84}Kr)} - qm_e)   + qm_e  + (R \cdot B{(^{84}Kr)}  - B{(^{84}Sr}))/c^2,
\end{equation}
Here, $M{(^{84}Sr)}$ and $M{(^{84}Kr)}$ represent the respective masses of the 
decay parent and daughter atoms
and $q$ denotes the charge state for singly charged ions ($q=1$). $m_{e}$ and $c$ correspond to the mass of an electron and the speed of light in vacuum. The electron binding energies, $B{(^{84}Sr)}$ and $B{(^{84}Kr)}$, are 5.69486745(12) eV and 13.9996055(20) eV, respectively, as obtained from~\cite{NIST_ASD}. 

The $Q$ value for the double-beta decay of $^{84}$Sr can be determined from the mass difference: $Q_{2\beta^+} = (M{(^{84}Sr)} - M{(^{84}Kr)})c^2$.



\begin{table}[!bht]
\scriptsize
\centering
\caption{The resulting $Q_{2\beta^+}$ and mass-excess values of $^{84}$Sr determined in this work based on the weighted mean of the cyclotron frequency ratio $\overline R$. The frequency ratio $\overline R$, $Q_{2\beta^+}$ values (in keV), and the mass excess (ME, in keV/c$^2$) of the parent, as determined in this work, are provided alongside the corresponding values from AME2020~\cite{Wang2021} for comparison.
 }
   \label{table:Q-value}    
\begin{tabular}{cccc}
\hline\noalign{\smallskip}
 & $\overline{R}$ & $Q_{2\beta^+}$ & ME \\
\hline\noalign{\smallskip}
This Work&  1.000 022 902 36(48) &  1790.115(37)& -80649.229(37) \\
AME2020 &  &  1789.8(12) & -80649.6(12) \\
\noalign{\smallskip}\hline\noalign{\smallskip}
\end{tabular}
\vspace*{1cm}  
\end{table}


\section{Results and discussion}

\subsection{Mass and $Q$-value determination}
A full scanning measurement (one cycle) of the magnetron phase, cyclotron phase, and center spot in sequence was completed in less than 3 minutes for each ion species of $^{84}$Kr$^{+}$ and $^{84}$Sr$^{+}$. In the analysis, the position of each spot 
within 5$\sigma$ of the ion distribution, 
was fit using the maximum likelihood method~\cite{Nesterenko2020,Ge2021-1}. 
Maximum-likelihood estimation with a Gaussian distribution was used for the parameter adjustment. Every four cycles were summed to ensure reasonable counts for fitting before determining the position of each spot. The phase angles were calculated accordingly based on the 
determined positions of the spots to deduce the cyclotron frequencies of each ion species.
The cyclotron frequency $\nu_{c}$ of the daughter ion $^{84}$Kr$^{+}$ was used as a reference and was linearly interpolated to the time of the measurement of the parent $^{84}$Sr$^{+}$ (ion of interest) to deduce the cyclotron frequency ratio $R$. Bunches with less than five detected ions per bunch were considered in the data analysis to reduce a possible cyclotron frequency shift due to ion-ion interactions~\cite{Kellerbauer2003, Roux2013}. Up to 5 detected ions per bunch were taken into acount for the analysis, and no count-rate related frequency shifts were observed in the analysis.
The temporal fluctuation of the magnetic field for JYFLTRAP was measured to be $\delta_B(\nu_{c})/\nu_{c} = \Delta t \times 2.01(25) \times 10^{-12}$/min~\cite{nesterenko2021}, where $\Delta t$ is the time interval between two consecutive reference measurements. The contribution of temporal fluctuations of the magnetic field to the final frequency ratio uncertainty was less than $10^{-10}$ since the $^{84}$Kr$^{+}$-$^{84}$Sr$^{+}$ measurements were interleaved with $\Delta t < 10$ minutes. 
To minimize the systematic uncertainty arising from the conversion of cyclotron motion to magnetron motion and potential distortion of the ion-motion projection onto the detector, the positions of the magnetron-motion and cyclotron-motion phase spots were deliberately chosen. The angle $\alpha_c$ between them was kept to less than 10 degrees, effectively reducing this uncertainty to a level well below 10$^{-10}$~\cite{Eliseev2014}.
Moreover, the commencement of the initial dipolar excitation with frequency $\nu_{+}$ was systematically scanned across one magnetron period (6 points), and the extraction was scanned over one cyclotron period (6 points) to mitigate any lingering effects of residual magnetron and cyclotron motion that might influence the distinct spots.
The measurements were conducted in eight separate time slots, each lasting around 4 hours
to ensure a consistent ion rate with a median value of 1-2 counts per bunch. For each slot of measurement, a weighted mean ratio $R_{4h}$ was calculated, and the maximum of internal and external errors~\cite{Birge1932} was selected. The final ratio $\overline R$ was then obtained as a weighted mean ratio of all $R_{4h}$ sets, taking into account the maximum of internal and external errors. 
%
The ions of the $^{84}$Kr$^{+}$ and $^{84}$Sr$^{+}$ were measured under similar conditions to minimize potential systematic shifts in the frequency ratio due to imperfections in the measurement trap. Mass-dependent systematic effects are negligible compared to the statistical uncertainty for mass doublets. No further systematic uncertainties were introduced, and these were confirmed through our previous measurements, as outlined in references  ~\cite{nesterenko2021,Nesterenko2020,Nesterenko22,Ramalho2022}.
The final frequency ratio $\overline{R}$ with its uncertainty, as well as the corresponding $Q_{2\beta^+}$ and mass-excess values, are 1.000 022 902 36(48), 1790.115(37) keV, and -80649.229(37) keV/c${^2}$ respectively.

\begin{figure}[!tbh]
\centering
\resizebox{0.49\textwidth}{!}{%
  \includegraphics{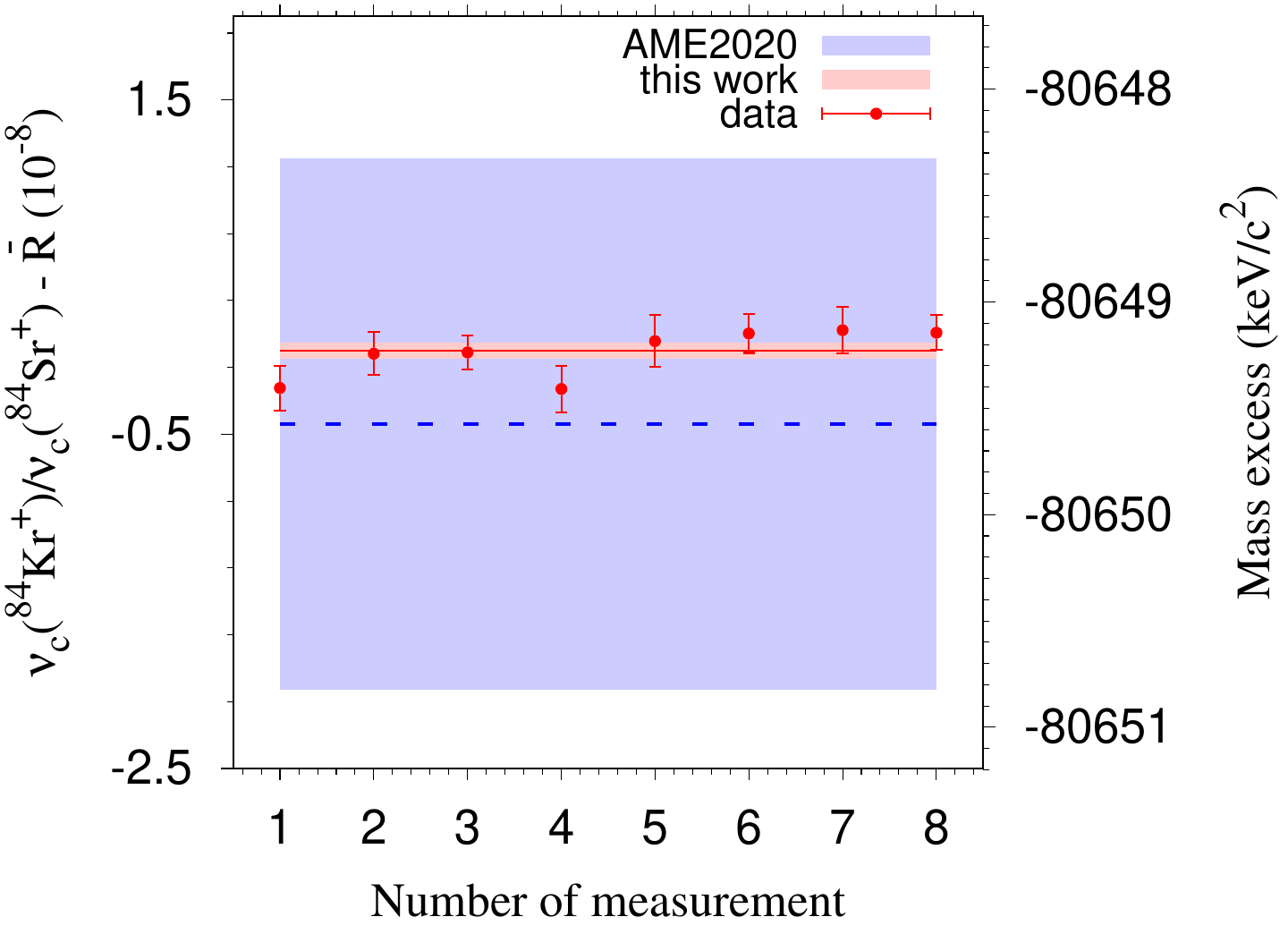}
  }
   \caption{(Color online). The deviation  (left axis) of the individually measured cyclotron frequency ratios $R_{4h}$ ($\nu_c$($^{84}$Kr$^{+}$)/$\nu_c$($^{84}$Sr$^{+}$)) from the final ratio value $\overline{R}$ and (right axis)) mass-excess values in this work compared to values adopted from AME2020~\cite{Huang2021,Wang2021}.    The red points, accompanied by uncertainties, represent individual data obtained with the PI-ICR method in eight distinct time slots. The solid red line illustrates the weighted average value from this work, $\overline{R}$ = 1.000 022 902 36(48), and its 1$\sigma$ uncertainty band is shaded in red. The dashed blue line represents the difference between the new value in this work and the one referred to in AME2020, with its 1$\sigma$ uncertainty area shaded in blue.}
   \label{fig:ratio}
\end{figure}

In Fig.\ref{fig:ratio}, the analysis results, which include all data sets, are compared to literature values. 
A comparison of these results  to the literature values are also tabulated in  Table.~\ref{table:Q-value}. 

The mass excess (-80649.229(37) keV/c${^2}$) and $Q_{2\beta^+}$ 
(1790.115(37) keV) 
from this work are both a factor of $\approx$30  more precise than those derived from the evaluated masses in AME2020~\cite{Huang2021,Wang2021}, but both agree well with the values documented in AME2020.
The  mass-excess value in AME2020 is derived primarily from two PTMS experiments~\cite{Keller07,Emma11} with an influence of 88.8\%. A slight contribution of 6.8\% is from endpoint energy measurements of $^{84}$Rb($\beta^-$)$^{84}$Sr~\cite{BOOIJ1971}  and the smallest influence of 2.1\% is related to a nuclear reaction experiment $^{84}$Sr(d,p)$^{85}$Sr ~\cite{MORTON1971}.

\begin{figure}[!tbh]
\centering
\resizebox{0.45\textwidth}{!}{%
  \includegraphics{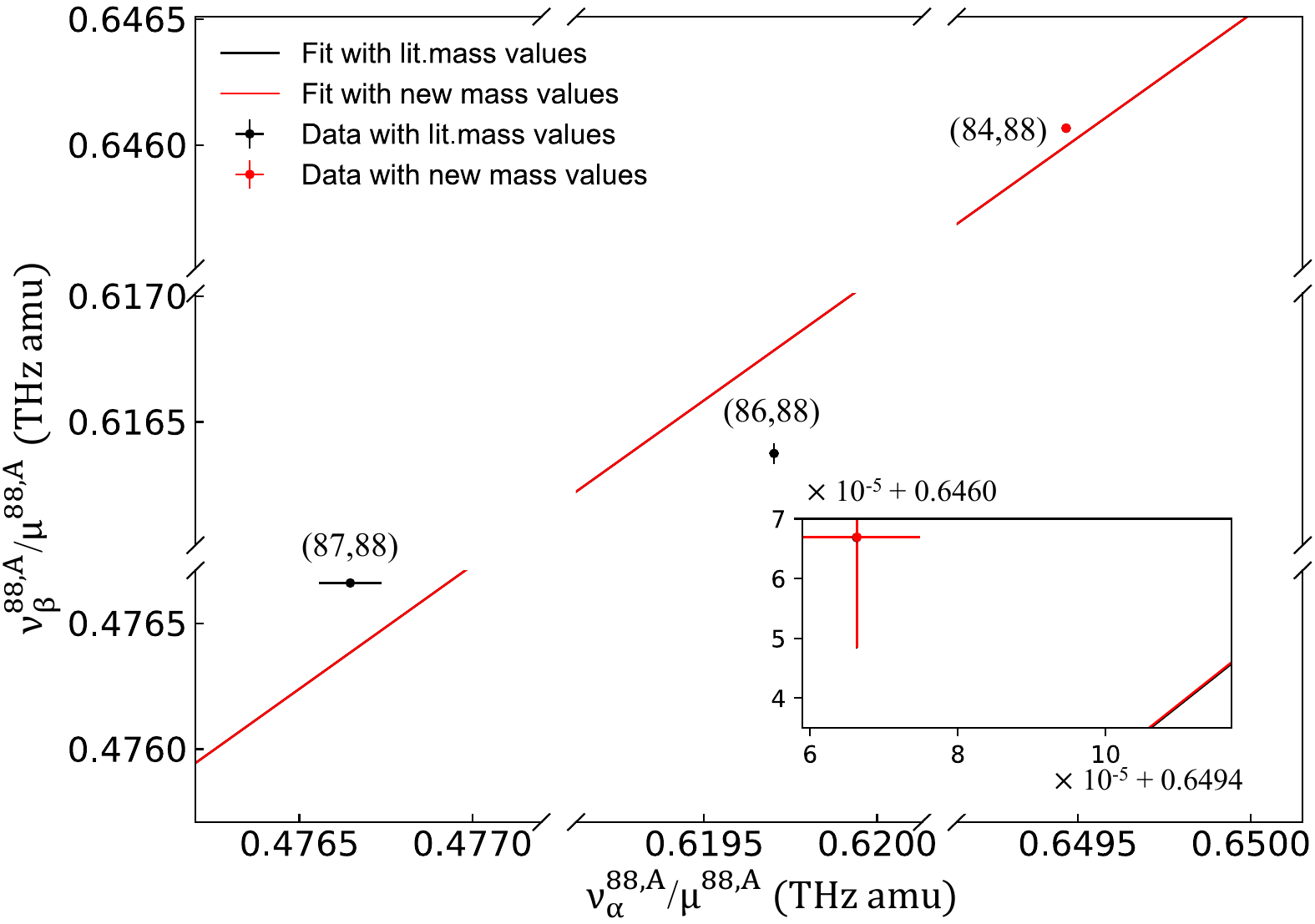}
}
   \caption{(Color online). King plot of the strontium isotope shifts in $\alpha$ (689 nm) and $\beta$ (698 nm) transitions as reported in ~\cite{miyake2019,Takano_2017}. The implication of the mass value of $^{84}$Sr from this work compared to values adopted from AME2020~\cite{Huang2021,Wang2021} 
can be observed.    
   Black data points and their accompanying error bars in black are obtained using the literature mass values, while the red data point with error bars is obtained with the new mass value from this work. Error bars and the difference between the $^{84}$Sr-$^{88}$Sr points derived from our measurement and from ~\cite{miyake2019,Takano_2017,Huang2021,Wang2021} are not visible at the scale being used due to a good agreement of this work with the AME2020. Linear fits were conducted for the set with literature data (black) and the new mass value (red), with a zoomed-in inset demonstrating a rather small change in the linear fit and no significant change in nonlinearity.}
      \label{fig:king}
\end{figure}

\subsection{King plot analysis}
The isotope shift between two isotopes with mass numbers $A_1$ and $A_2$, ($\nu ^{A_1A_2}_i$) is defined as the difference in transition frequencies: $\nu ^{A_1A_2}_i=\nu ^{A_1}_i - \nu ^{A_2}_i$. The primary contributions to the isotope shift arise from the mass shift (MS) and field shift (FS).
The MS results from the mass difference of isotopes $A_1$ and $A_2$ and is expressed as an electronic coefficient $k_i$ multiplied by the isotope-dependent 
inverse mass factor, given by: $\mu^{A_1A_2} = 1/m_{A_2}-1/m_{A_1}$.
The FS originates from the differing volumes of the two isotopes, factorized into an electronic, isotope-independent coefficient $F_i$ and the charge radius variance $\delta \langle r^2\rangle_{A_1A_2} = \langle r^2_{A_1}\rangle - \langle r^2_{A_2}\rangle$. 
The isotope shift composition of a transition $i$ in terms of electronic and nuclear quantities factorizes into the total isotope shift equation: $\nu ^{A_1A_2}_i = k_i \mu^{A_1A_2} + F_i \delta \langle r^2\rangle_{A_1A_2} + ...$, where the first term represents the MS, and the second term represents the FS~\cite{king1984isotope,king63}, with the dots denoting potential higher-order corrections and new physics contributions. 
The frequency shifts are commonly normalized by the 
inverse mass factor $\mu^{A_1A_2}$ to obtain the modified isotope shift, $\nu^{A_1A_2}_i/\mu^{A_1A_2}$~\cite{Viatkina23}. Consequently, the MS (the sum of the normal MS and specific MS) reduces to the electronic factor $k_i$, while the FS factor $F_i$ is multiplied by the modified charge radius variance, 
$\delta \langle r^2\rangle_{A_1A_2}/\mu^{A_1A_2}$, establishing a linear dependence between the two sets of modified frequency shifts known as the King linearity~\cite{king63}. 
If the isotope shifts are measured for more than one transition ($i$ and $j$), one could eliminate the typically poorly known difference of the mean squared nuclear charge radii $\delta \langle r^2\rangle_{A_1A_2}$  and to write the so-called King relation~\cite{king63}: 
${\nu_i^{A_1A_2}}/{\mu^{A_1A_2}} = K_i  -  {F_i}/{F_j}\cdot K_j  + {F_i}/{F_j} \cdot {\nu_j^{A_1A_2}}/{\mu^{A_1A_2}}$.
To quantify the observed linearity, a measure of nonlinearity is defined~\cite{flam18,Berengut2018}. A King plot analysis can be used to systematically quantify and visually examine isotope shifts in various atomic transitions referenced to the same isotope.

The refined mass-excess value obtained for $^{84}$Sr was employed to carry out an updated King plot analysis for two transitions at  
689 nm and 698 nm, 
as detailed in~\cite{miyake2019,Takano_2017}.
Figure~\ref{fig:king} displays the King plot featuring both the literature mass values from AME2020 and our updated mass value. The main uncertainties in the plot are from the frequency shifts measurements in the transitions between $^{84}$Sr, $^{86}$Sr, and $^{87}$Sr relative to $^{88}$Sr as adopted from~\cite{miyake2019}. The transition of $^{87}$Sr-$^{88}$Sr at 698 nm is from~\cite{Takano_2017} which has a better precision than that of ~\cite{miyake2019}. 
The  observed nonlinearity from~\cite{miyake2019},  achieved through our new mass value for $^{84}$Sr is evident, and 
the reduced mass uncertainty will enhance the possible precision of isotope-shift data, overcoming previous limitations imposed by $^{84}$Sr.
Given that the uncertainty in the mass contributes to the calculation of the modified isotope shifts in similar ways, the impact on the linearity of the King plot is preserved due to the agreement of our result with the literature value. 

{A linear orthogonal distance regression analysis~\cite{ODR2004} that provides unified standard error estimates for the uncertainties implanted in $x$ and $y$ directions, was conducted to determine the slope and intercept. The resulting slope and intercept from the fit to the AME2020 mass values and the updated mass value are 0.98148(51) and 8578(326) MHz$\cdot$amu, respectively.} 
Our results align well with the fitting values of 0.981(5) and 8560(3450) MHz$\cdot$amu, reported in~\cite{miyake2019}, offering smaller uncertainties. These results confirm the nonlinearity announced in~\cite{miyake2019}, using the nonlinearity measure defined in ~\cite{flam18,Berengut2018}.
The uncertainty from the mass values now allows for potential future 
precision measurements of atomic transitions using optical spectroscopy to reach a relative precision by more than three orders of magnitude, thus potentially leading to an unprecedented sensitivity for new physics. 
This will achieve the spectroscopic precision necessary to attain new limits on a spin-independent fifth force interaction, as current calculations place the required uncertainty at < 1 Hz~\cite{Ludlow15,Berengut2018}. Such experiments require a measurement of the optical atomic transitions with a relative precision of 10$^{-15}$ or lower, a challenging accomplishment feasible currently at a few optical clock laboratories~\cite{Ludlow15}. 




\section{Conclusion and outlook}
In summary, we have performed direct high-precision mass and double-beta decay $Q$ value measurements of $^{84}$Sr using the PI-ICR technique with the JYFLTRAP double PTMS. 
The newly refined mass and $Q$ values for $^{84}$Sr were in agreement with the adopted values in AME2020, while a precision approximately 30 times higher than those adopted in AME2020 were achieved.
The new mass value was utilized to conduct an updated King plot analysis for transitions at 
689 nm and 698 nm 
for $^{84,86,87}$Sr in relation to $^{88}$Sr.  
Our results reveal a nonlinear King plot, consistent with prior research and featuring reduced uncertainties.
The current contribution of the mass uncertainty to the King plot analysis enables future precision enhancements by several orders of magnitude in optical spectroscopy, mitigating statistical and systematic errors in both transitions. 
This potential advancement in both transitions could lead to unparalleled sensitivity to new physics.
More precise isotope shift measurements of Sr isotopes, providing stringent experimental constraints on King linearity, are highly required. Performing such a measurement with a radioactive isotope e.g., $^{90}$Sr, to avoid complications due to hyperfine structure, is feasible at the IGISOL facility. In conjunction with additional measurements of isotope shifts in elements such as calcium, 
this will contribute to testing the predictions from atomic theory 
and imposing constraints on new physics.

\begin{acknowledgement}

We acknowledge the staff of the Accelerator Laboratory of University of Jyv\"askyl\"a (JYFL-ACCLAB) for providing stable online beam. We thank the support by the Academy of Finland under the Finnish Centre of Excellence Programme 2012-2017 (Nuclear and Accelerator Based Physics Research at JYFL) and projects No. 306980, No. 312544, No. 275389, No. 284516, No. 295207, No. 314733, No. 315179, No. 327629, No. 320062, No. 354589 and 345869. The support by the EU Horizon 2020 research and innovation program under grant No. 771036 (ERC CoG MAIDEN) is acknowledged.  
We express gratitude for the productive results stemming from conversations with R. de Groote and X. F. Yang.


\end{acknowledgement}

%
 \bibliographystyle{unsrt} 
 \bibliography{my-final-bib-from-jabref.bib}
%
%
%

\end{document}